\documentclass[%
 reprint,
superscriptaddress,
 amsmath,amssymb,
 aps,
]{revtex4-2}

\usepackage{graphicx}% Include figure files
\usepackage{dcolumn}% Align table columns on decimal point
\usepackage{bm}% bold math

\usepackage[version=3]{mhchem} % Formula subscripts using \ce{}
\usepackage{amsfonts}
\usepackage{xcolor}
\usepackage{soul}

\begin{document}

\title{Optical excitation of multiple standing spin modes in 3D optomagnonic nanocavities}
\preprint{APS/123-QED}

\author{Daria O. Ignatyeva}
\email{daria.ignatyeva@gmail.com}
\affiliation{Russian Quantum Center, 121205 Moscow, Russia}
\affiliation{Faculty of Physics, M.V. Lomonosov Moscow State University, 119991 Moscow, Russia}

\author{Denis M. Krichevsky}
\affiliation{Russian Quantum Center, 121205 Moscow, Russia}
\affiliation{Moscow Institute of Physics and Technology, Moscow, Russia}

\author{Dolendra Karki}
\affiliation{Physics Department, Michigan Technological University, Houghton, Michigan, U.S.A.}

\author{Anton Kolosvetov}
\affiliation{Russian Quantum Center, 121205 Moscow, Russia}
\affiliation{Center for Photonics and 2D Materials, Moscow Institute of Physics and Technology (National Research University), Moscow, Russia}

\author{Polina E. Zimnyakova}
\affiliation{Russian Quantum Center, 121205 Moscow, Russia}
\affiliation{Moscow Institute of Physics and Technology, Moscow, Russia}

\author{Alexander N. Shaposhnikov}
\affiliation{Institute of Physics and Technology, V.I. Vernadsky Crimean Federal University, 295007 Simferopol, Russia}

\author{Vladimir N. Berzhansky}
\affiliation{Institute of Physics and Technology, V.I. Vernadsky Crimean Federal University, 295007 Simferopol, Russia}

\author{Miguel Levy}
\affiliation{Physics Department, Michigan Technological University, Houghton, Michigan, U.S.A.}

\author{Alexander I. Chernov}
\affiliation{Russian Quantum Center, 121205 Moscow, Russia}
\affiliation{Center for Photonics and 2D Materials, Moscow Institute of Physics and Technology (National Research University), Moscow, Russia}

\author{Vladimir I. Belotelov}
\affiliation{Faculty of Physics, M.V. Lomonosov Moscow State University, 119991 Moscow, Russia}
\affiliation{Russian Quantum Center, 121205 Moscow, Russia}
%\footnote{}

%\begin{tocentry}
%\includegraphics[width=3.3in]{TOC.png}
%\end{tocentry}

\date{\today}

\begin{abstract}

We report the first experimental observation of multiple standing spin modes in 3D optomagnonic nanocavity formed by nanometer-sized iron-garnet nanocylinder. We show that launching of standing spin modes is achieved due to a high confinement of the optically generated effective magnetic field caused by the localized optical resonance. Quantization and spin-wave mode inhomogeneity is achieved in each of the three spatial dimensions. The presented approach opens new horizons of 3D optomagnonics by combining nanophotonic and magnonic functionalities within a single nanocavity.
\end{abstract}

\maketitle

%\section*{Introdution}
Optomagnetism~\cite{kimel20222022,kalashnikova2015ultrafast, kimel2007femtosecond} is important for ultrafast, local, and reconfigurable spin control. This opens novel opportunities for fast and energy-efficient information processing~\cite{kajiwara2010transmission,chen2021reconfigurable}, including quantum information technologies~\cite{li2021quantum,lachance2019hybrid}, logical elements~\cite{kolosvetov2022concept,jamali2013spin}, memory~\cite{kimel2019writing,stanciu2007all,gundougan2015solid,stupakiewicz2019selection,stupakiewicz2017ultrafast}, dynamic holography~\cite{makowski2022dynamic},  and other spin-based devices~\cite{nikitov2015magnonics}. Optomagnetism is based on the action of the femtosecond laser pulse on the spin system of a magnetic material, which may lead to different scenarios of spin dynamics, including the excitation of spin precession and spin waves~\cite{satoh2012directional,freeman2012all,savochkin2017generation,filatov2022spectrum,gerevenkov2023unidirectional} and magnetization reversal~\cite{stanciu2007all,stupakiewicz2017ultrafast,stupakiewicz2019selection,ignatyeva2021magnetization}.

Miniaturization is one of the key problems in optomagnetism due to a diffraction limit. To overcome this limit and excite spin dynamics at the nanoscale, various nanoplasmonic~\cite{bossini2016magnetoplasmonics,maccaferri2019coupling,dutta2017surface,zimnyakova2022plasmonic,ignatyeva2019plasmonic,yang2023inverse} and dielectric magnetophotonic~\cite{qin2022nanophotonic,ignatyeva2022all,chernov2020all,sylgacheva2022spatially,krichevsky2021selective} structures may be used to localize light at the spots of subwavelength dimensions. At the same time, the limited size of the optomagnetic cavity imposes selection rules on the spin wavevector and results in its quantization and the appearance of a standing spin modes set. While for micrometer-sized objects the magnetostatic splitting of such modes eigenfrequencies is small~\cite{edwards2013magnetostatic}, the situation drastically changes with the further miniaturization of the device size down to nanometer scales. The considerable splitting of the frequencies of different spatial eigenmodes arises in this case due to the interplay of the magnetodipole and exchange interactions~\cite{chernov2020all,krichevsky2021selective}. As a result, magnets smaller than a micron could be thought of as optomagnonic nanocavities supporting both localized optical and spin modes.

The optical impact on spins can be described in terms of an effective magnetic field induced by light inside a magnet~\cite{kalashnikova2015ultrafast}. Localization and the direction of the optically-induced effective magnetic field could be easily tuned in the nanostructure by varying the parameters of the laser pulse. It allows one to selectively excite the desired modes~\cite{chernov2020all,krichevsky2021selective}. This is an important advantage of the optomagnetic means compared to the traditional ones using a ferromagnetic resonance technique and microwave antennas~\cite{kakazei2004spin,polylakh2021FMR,li2019nutation}. Apart from that, addressing a single nanomagnet via a microscopic nanoantenna requires its precise positioning, for example via the movable substrate~\cite{dobrov2020nanosc}, increasing the complexity of the setup and likely limiting the operation rate.

Recently demonstrated high-order spin modes excited optomagnetically by laser pulses in smooth films~\cite{deb2019femtosecond,deb2022controlling} and nanophotonic structures~\cite{chernov2020all,krichevsky2021selective} are the standing spin waves across the film thickness, while in the lateral direction they still possess quasi-homogeneous profiles.

In the present paper, we take the next step forward and report optical excitation of standing spin modes in optomagnonic nanocavities confined in all three spatial dimensions. We demonstrate that a femtosecond laser pulse in a magnetic nanocylinder acts like a point-like source of magnetization dynamics due to the excited localized optical modes of the nanocylinder providing a highly confined effective magnetic field. Due to this, a set of nanocavity spin modes of different orders with different frequencies is launched.

%\section*{Results}
Optical excitation of the confined spin waves is performed by illuminating with femtosecond laser pulses on an array of the bismuth-substituted iron-garnet nanocylinders. The nanocylinders are 450 nm in diameter and 515 nm in height. They are periodically arranged in a 2D square lattice with a period $P$=900 nm (see Supplemental Material for the fabrication details). The large separation between the cylinders makes both optical~\cite{zimnyakova2021two} and magnetodipole interactions between them negligible. Thus, both the optical and spin mode properties of the system are mostly determined by the properties of an individual nanocylinder.

\begin{figure*}[htbp]
\centering
(a)~~~~~~~~~~~~~~~~~~~~~~~~~~~~~~~~~~(b)~~~~~~~~~~~~~~~~~~~~~~~~~~~~~~~~~~(c) \\
\includegraphics[width=0.32\linewidth]{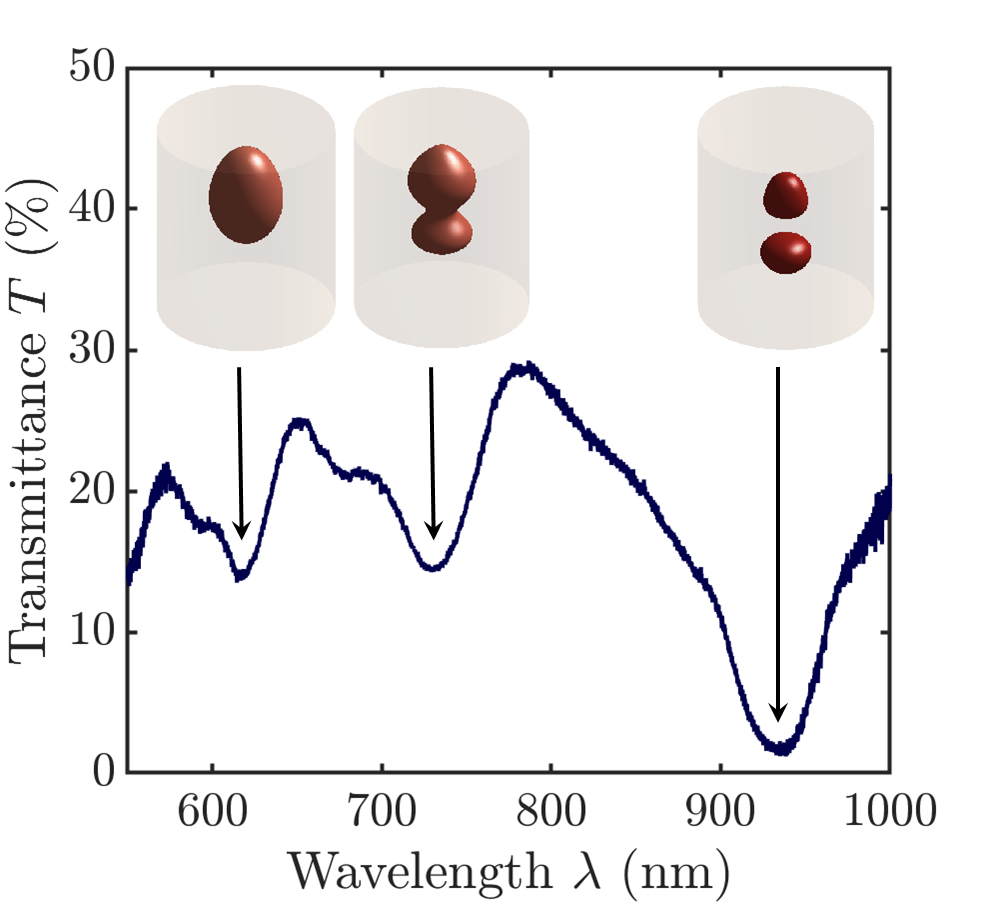} 
\includegraphics[width=0.32\linewidth]{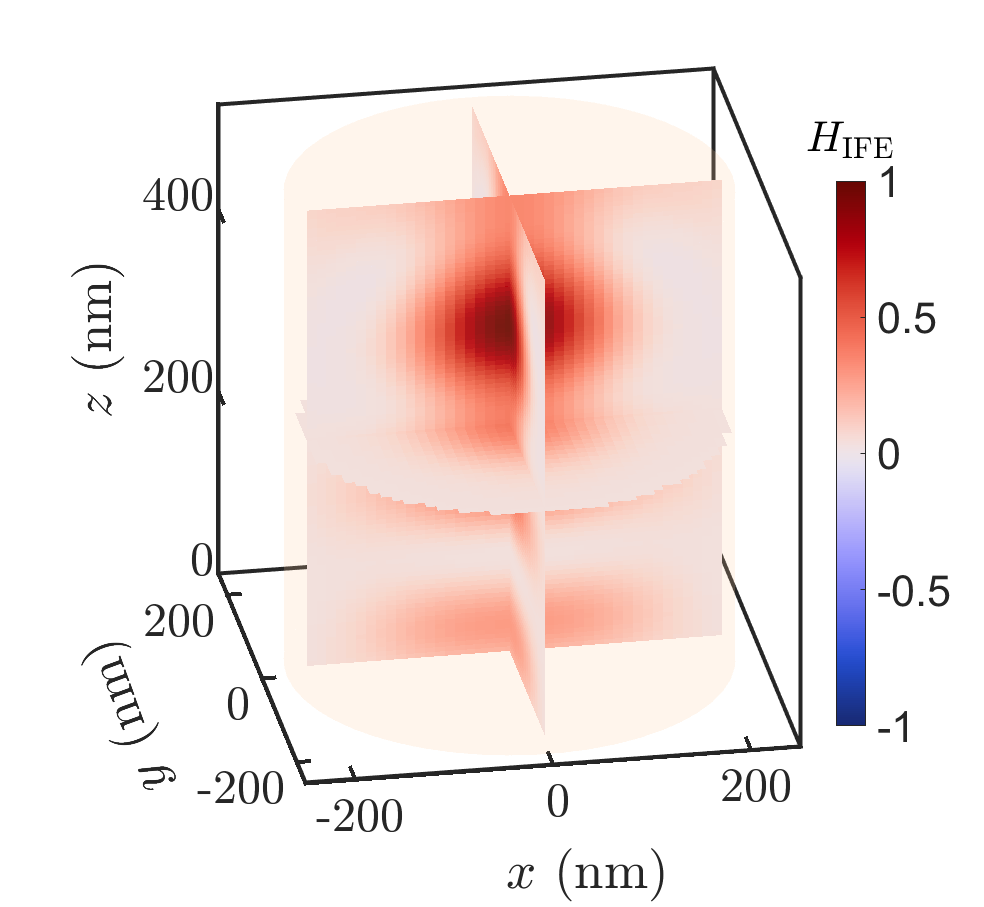} 
\includegraphics[width=0.32\linewidth]{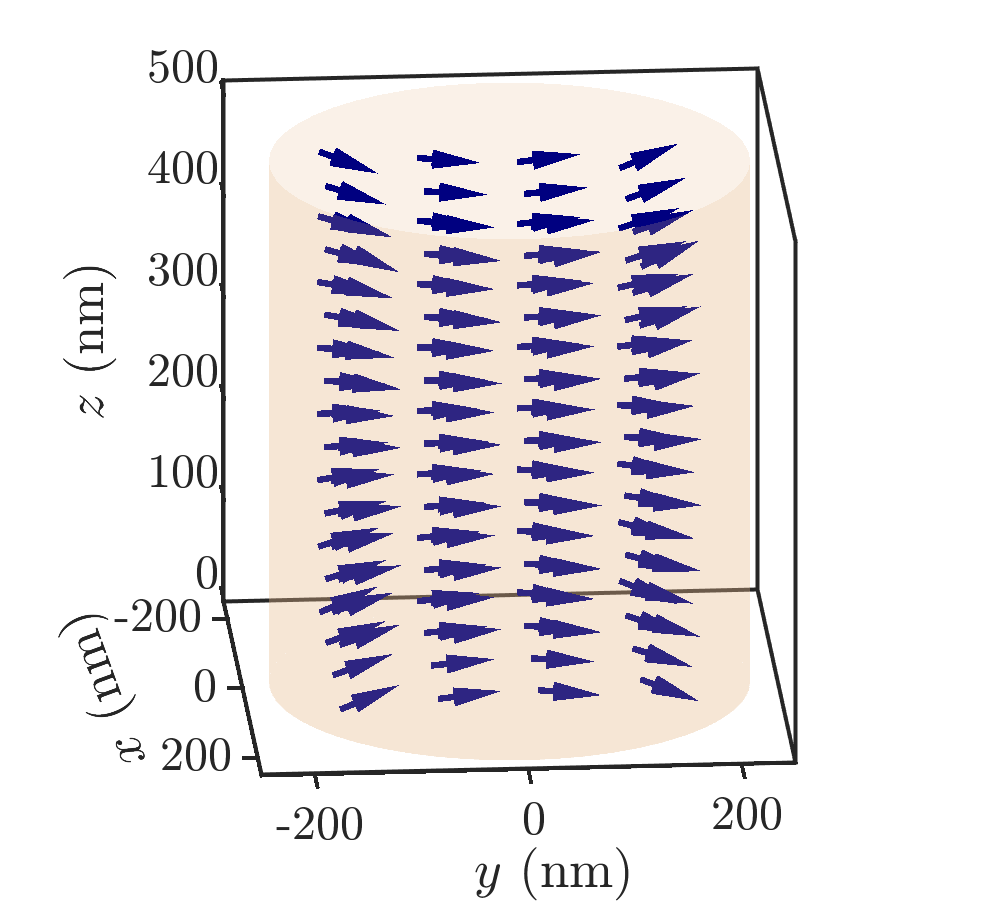} 
\caption{Optical excitation of spin modes. (a) Experimental transmittance spectra of the nanocylinder array. Insets represent the regions inside the nanocylinder where the $z$-component of the $H_\mathrm{IFE}$ field is maximal, shown as the isosurfaces at $e^{-1}$ level. (b) The distribution of the normalized $H_\mathrm{IFE}$ field $z$-component inside a nanocylinder calculated at $\lambda=730$~nm wavelength shown at three central cross-sections of a cylinder. (c) Nanocylinder static magnetization under external magnetic field $H_\mathrm{ext}=H_y$ applied perpendicular to its axis.}
\label{fig: Optical}
\end{figure*}

The transmittance spectrum (Fig.~\ref{fig: Optical}a) of a nanocylinder exhibits a set of optical resonances excited in a dielectric nanocylinder. These resonances may be described as the Fabry–Pérot resonances of a guided modes of the cylindrical waveguides~\cite{zimnyakova2021two}. The electromagnetic field of the light is strongly localized inside the nanocylinder. Consequently, circular polarization of the incident pulse induces the effective magnetic field via the inverse Faraday effect~\cite{kalashnikova2015ultrafast} (IFE) $\mathbf{H}_\mathrm{IFE}\propto\mathrm{Im}\left[\mathbf{E}\times\mathbf{E}^*\right]$ directed predominantly along $z$ axis, which is also localized within the nanocylinder, as shown in the insets of Fig.~\ref{fig: Optical}a and in Fig.~\ref{fig: Optical}b.

For optical excitation of the spin modes, we choose the resonance at the wavelength $\lambda$=730~nm. This wavelength is in the spectral range optimal for the magneto-optical figure of merit, which is due to a sort of trade-off between the high gyration values and low damping of the iron-garnet material itself. It allows one to perform optical pumping quite efficiently and to avoid thermal effects. Since at this wavelength iron-garnets are rather transparent, the main impact of the circularly polarized pulses on spins is due to the inverse Faraday effect~\cite{kalashnikova2015ultrafast}. Figure~\ref{fig: Optical}b shows a strong confinement of the effective magnetic field $H_\mathrm{IFE}$ generated by a laser pulse via the inverse Faraday effect (IFE) in the central part of a cylinder. While the volume of the nanocylinder itself is $V_\mathrm{cyl}=8\cdot10^{-2}\mu\mathrm{m}^3$, the effective volume of laser-induced IFE field calculated as $V_\mathrm{IFE}=\int{H_\mathrm{IFE}(\mathbf{r})\mathrm{d}V}/\max|H_\mathrm{IFE}|$ is an order lower $V_\mathrm{IFE}=7.6\cdot10^{-3}\mu\mathrm{m}^3$. This causes the optical pump to act as a point-like source, exciting multiple standing spin modes.

Nanocylinders were magnetized up to saturation by an in-plane magnetic field $\mathbf{H}_\mathrm{ext}$ applied to the sample in-plane, perpendicular to the axes of the cylinders. It is important that the height $H$ and diameter $D$ of the cylinder are close to each other: $H\approx D$, so that the aspect ratio of the nanocavity is close to 1:1. This makes the contribution of the demagnetizing fields quite prominent. The resulting magnetization direction inside the cylinder is inhomogeneous (Fig.~\ref{fig: Optical}c). A significant tilt of the static magnetization from the $\mathbf{H}_\mathrm{ext}$ direction, an increase of the normal, and a decrease of the in-plane magnetization components emerge near the cylinder walls. Such a peculiar magnetization distribution makes the frequencies of the modes sensitive to their localization and amplitude distribution inside the nanocavity. The difference in spin mode frequencies excited in a nanocavity is important since it allows us to identify and study the modes separately.

The effective magnetic field generated by the femtosecond laser pulse via the IFE deflects the magnetization from its equilibrium value and thus launches the spin dynamics. The high confinement of this field makes optical excitation similar to a point-like source and thus makes it possible to excite a set of the standing spin modes with inhomogeneous and sign-changing profiles. The frequencies of these modes are different from each other due to the following reasons. First of all, according to the theory (Fig.~\ref{fig: Optical}c) of demagnetizing fields, the magnetization of the cylinder is non-uniform, and in-plane components decrease near the cylinder walls. Thus, the frequencies of the modes localized near the cylinder walls would be smaller than the ones localized at the cylinder center. Apart from that, the more inhomogeneous a mode profile is, the more significant the exchange field contribution is, and the higher the spin mode frequency is.

\begin{figure}[htbp]
\centering
(a)~~~~~~~~~~~~~~~~~~~~~~~~~~~~~~~~~~~~~~~~~~~~~~~~~~~~~~~(b) \\
\includegraphics[width=0.48\linewidth]{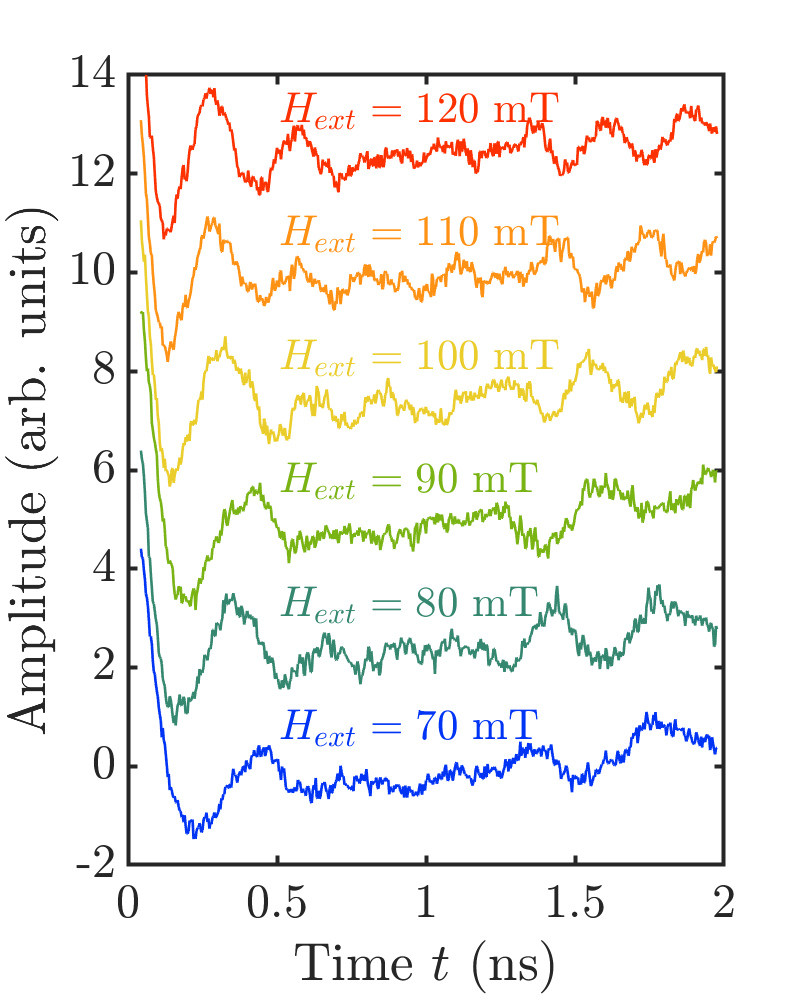} ~\includegraphics[width=0.48\linewidth]{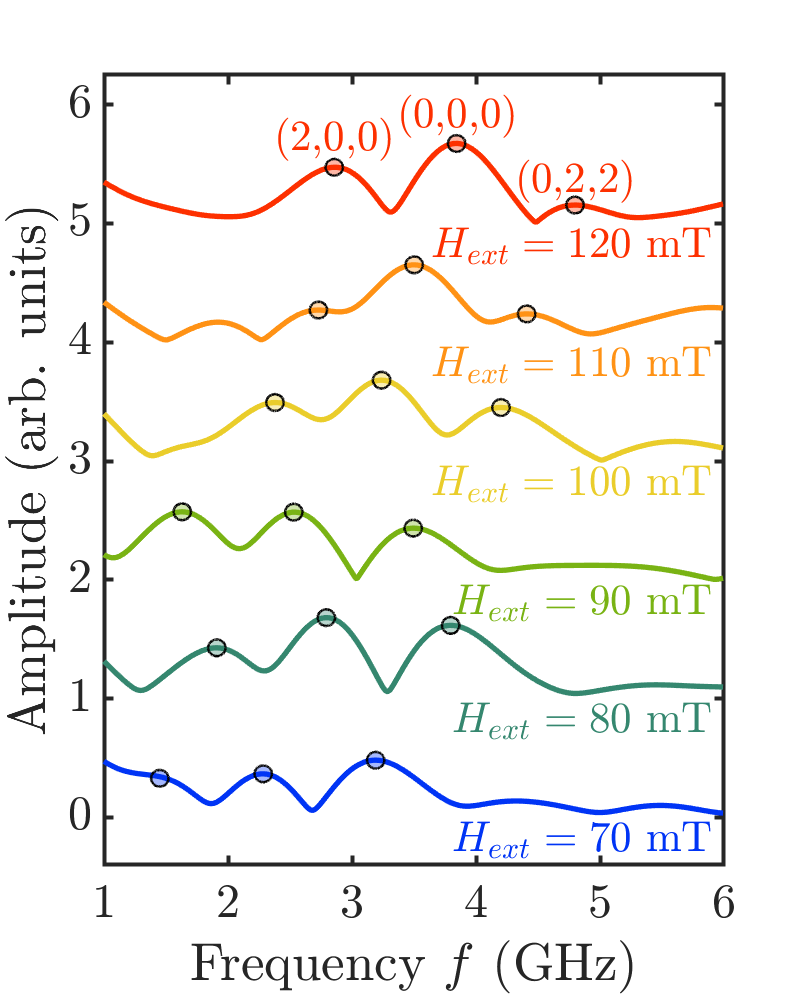} \\
(c)~~~~~~~~~~~~~~~~~~~~~~~~~~~~~~~~~~~~~~~~~~~~~~~~~~~~~~~(d) \\
\includegraphics[width=0.47\linewidth]{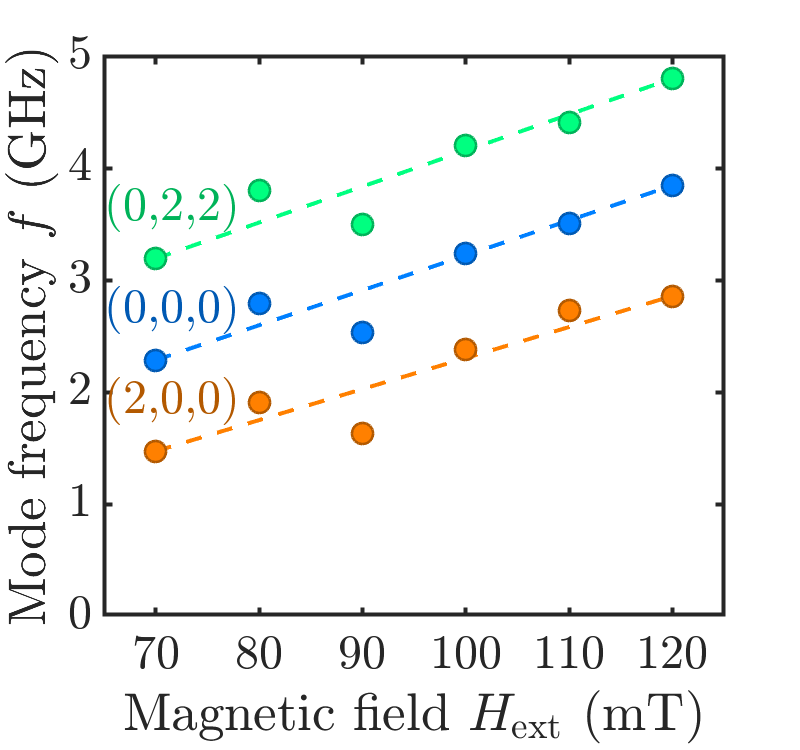} 
\includegraphics[width=0.47\linewidth]{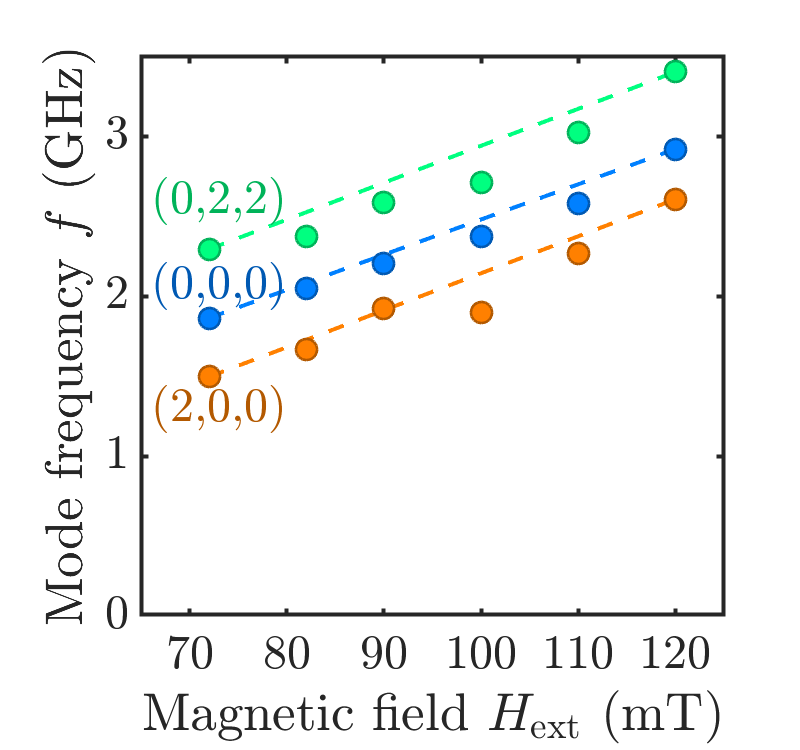} 
\caption{Spin dynamics in the nanocylinder excited by a femtosecond pulse at $\lambda=730$~nm. (a) Magnetization precession and (b) its Fourier spectra measured by the transient Faraday rotation of the probe pulse. The peaks corresponding to the excited standing spin modes are marked as $(n_x, n_y, n_z)$ according to the number of nodes along a certain direction (see the details in the text). (c,d) Magnetic field dependence of the frequencies of standing spin modes obtained (c) experimentally and (d) numerically.}
\label{fig: Precession}
\end{figure}

Simultaneous excitation of multiple spin modes results in a complex beating pattern of the temporal dependence of the magnetization precession itself (Fig.~\ref{fig: Precession}a). The Fourier spectra show several distinct peaks corresponding to the eigenfrequencies of the excited modes (Fig.~\ref{fig: Precession}b). A detailed analysis of the excited spin mode profiles with the corresponding frequencies is provided below. Hereafter, we mark these peaks as $(n_x, n_y, n_z)$ according to the nature of the standing spin modes and the corresponding number of nodes along the $x,~y,~z$ spatial directions.

Eigenfrequencies of the spin modes depend on the external magnetic field $H_\mathrm{ext}$ applied to the nanocylinder. Experimental results (Fig.~\ref{fig: Precession}c) show a good qualitative correspondence with the numerical simulations (Fig.~\ref{fig: Precession}d), while some quantitative differences might be caused by inaccuracies in nanocylinder fabrication processes. The numerical simulations of spin dynamics were performed using mumax3 software~\cite{vansteenkiste2014design} (see Supplemental Material for details). At the same time, the analytical description of the nanocylinder eigenmodes is quite complicated. On the one hand, as the ratio of height and diameter of the nanocavity is nearly 1:1, there are 3 degrees of freedom. On the other hand, demagnetizing fields are significant, and the axial symmetry is broken by the external magnetic field $H_\mathrm{ext}$ applied perpendicular to the nanocylinder axis.  A good agreement between the experimental results and numerical simulations makes it possible to perform a numerical analysis of the profiles and types of spin modes excited in a nanocavity.

\begin{figure*}[htbp]
\centering(a)~~~~~~~~~~~~~~~~~~~~~~~~~~~~~~~~~~~(b)~~~~~~~~~~~~~~~~~~~~~~~~~~~~~~~~~~~(c)\\~\\
\includegraphics[width=0.32\linewidth]{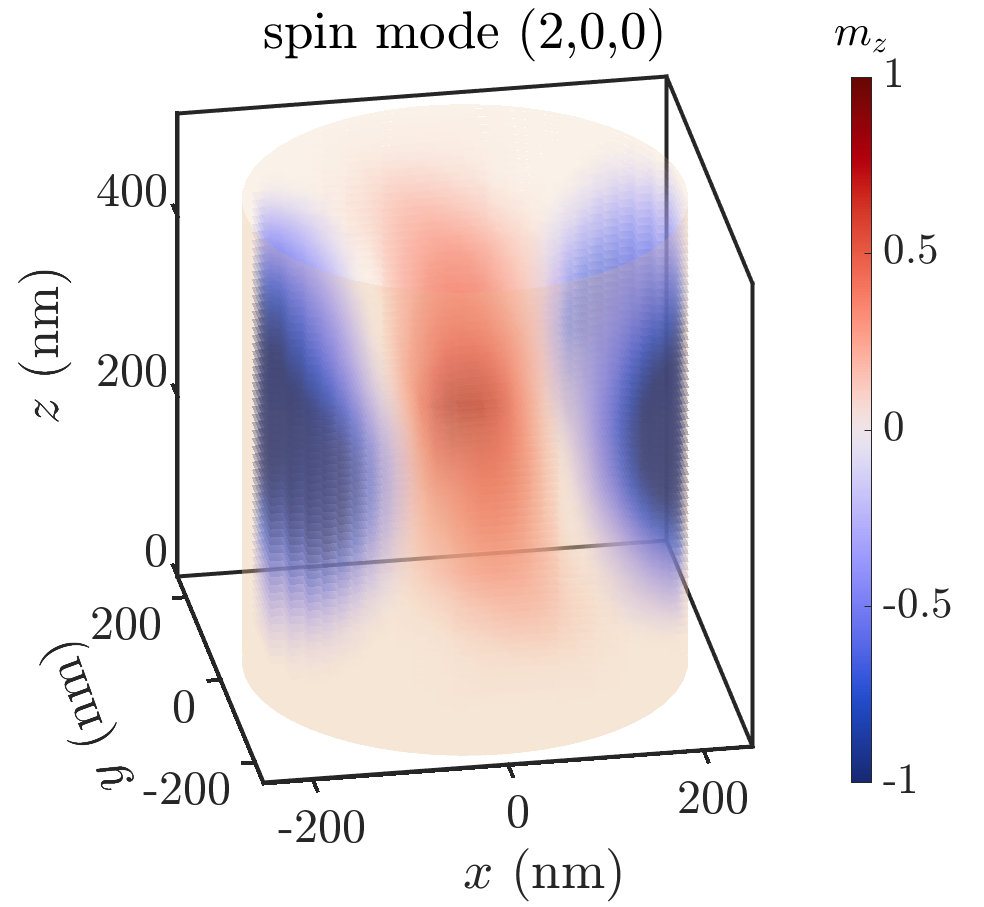}~~~
\includegraphics[width=0.32\linewidth]{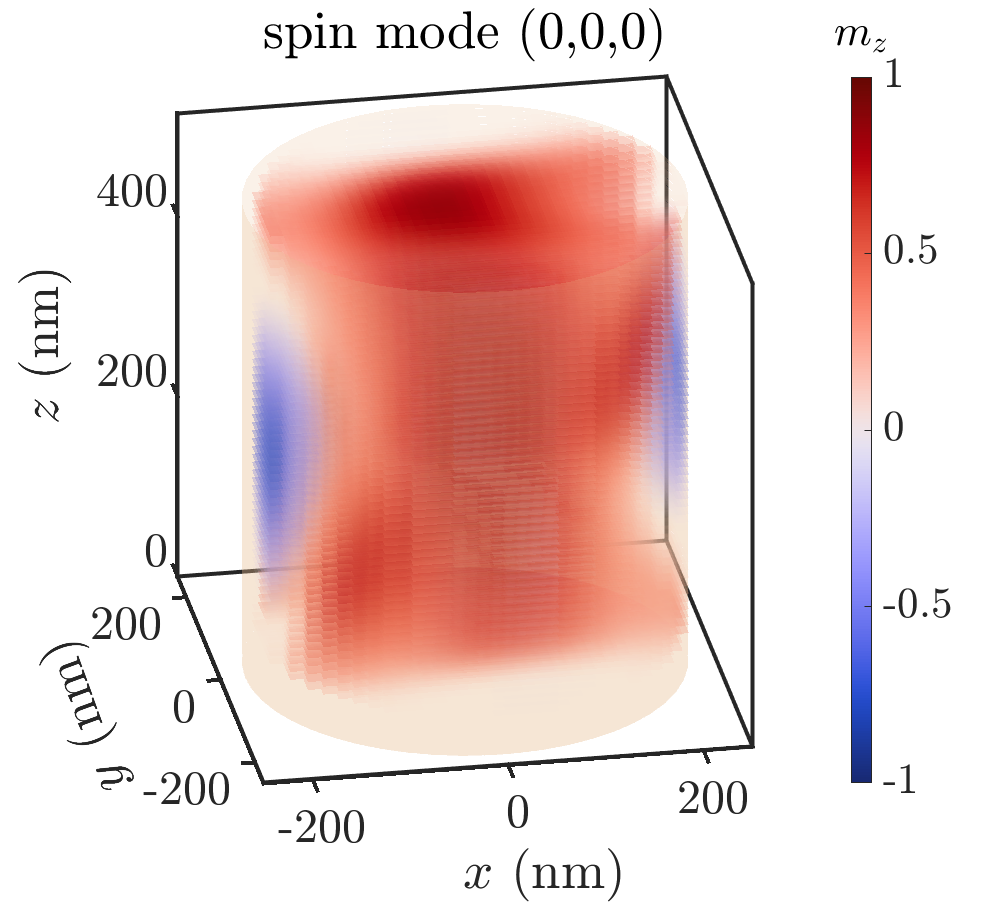}~~~
\includegraphics[width=0.32\linewidth]{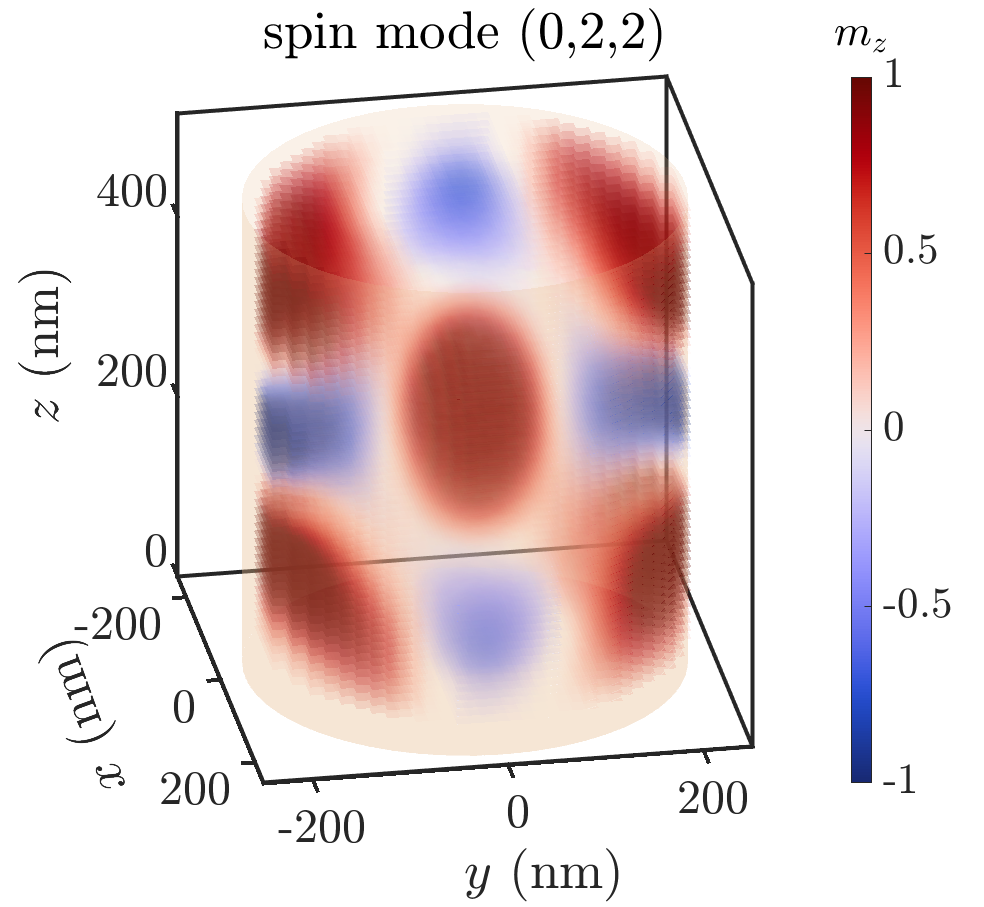}\\~\\
(d)~~~~~~~~~~~~~~~~~~~~~~~~~~~~~~~~~~~(e)~~~~~~~~~~~~~~~~~~~~~~~~~~~~~~~~~~~(f)\\~\\
\includegraphics[width=0.32\linewidth]{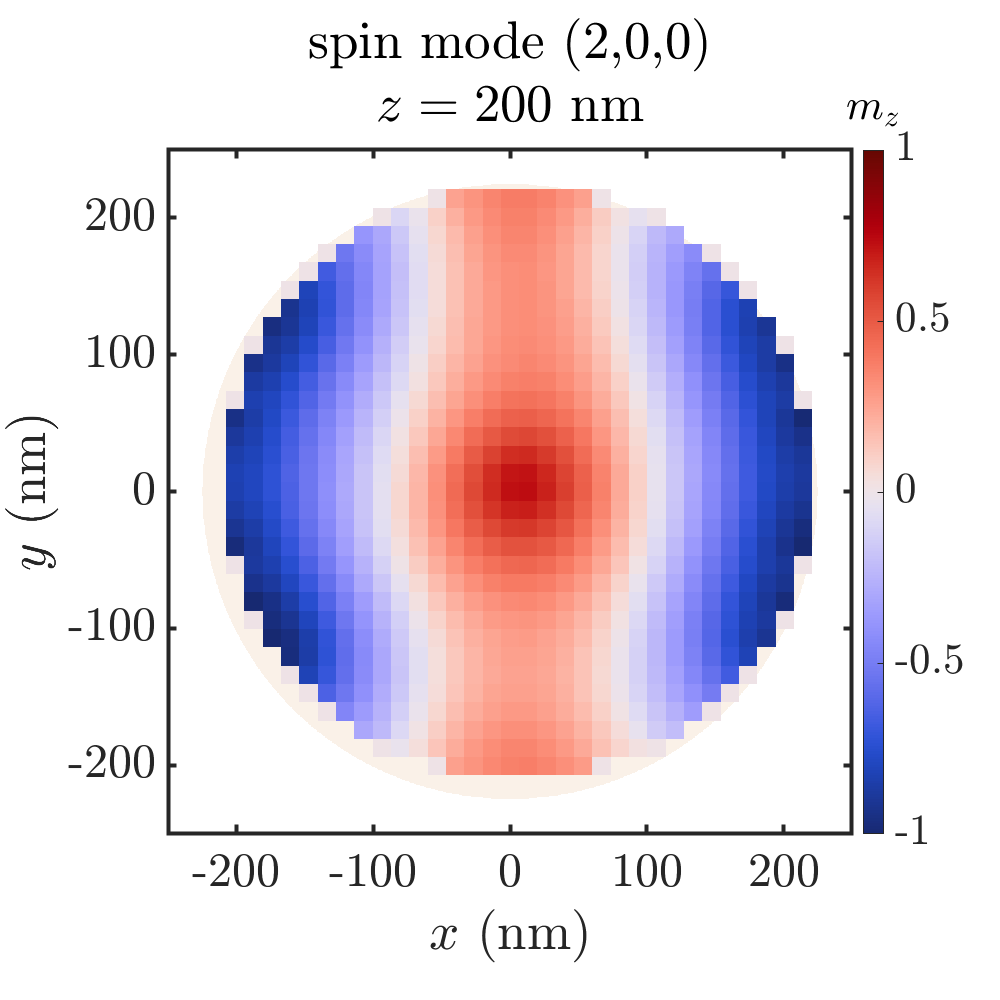}~~~
\includegraphics[width=0.32\linewidth]{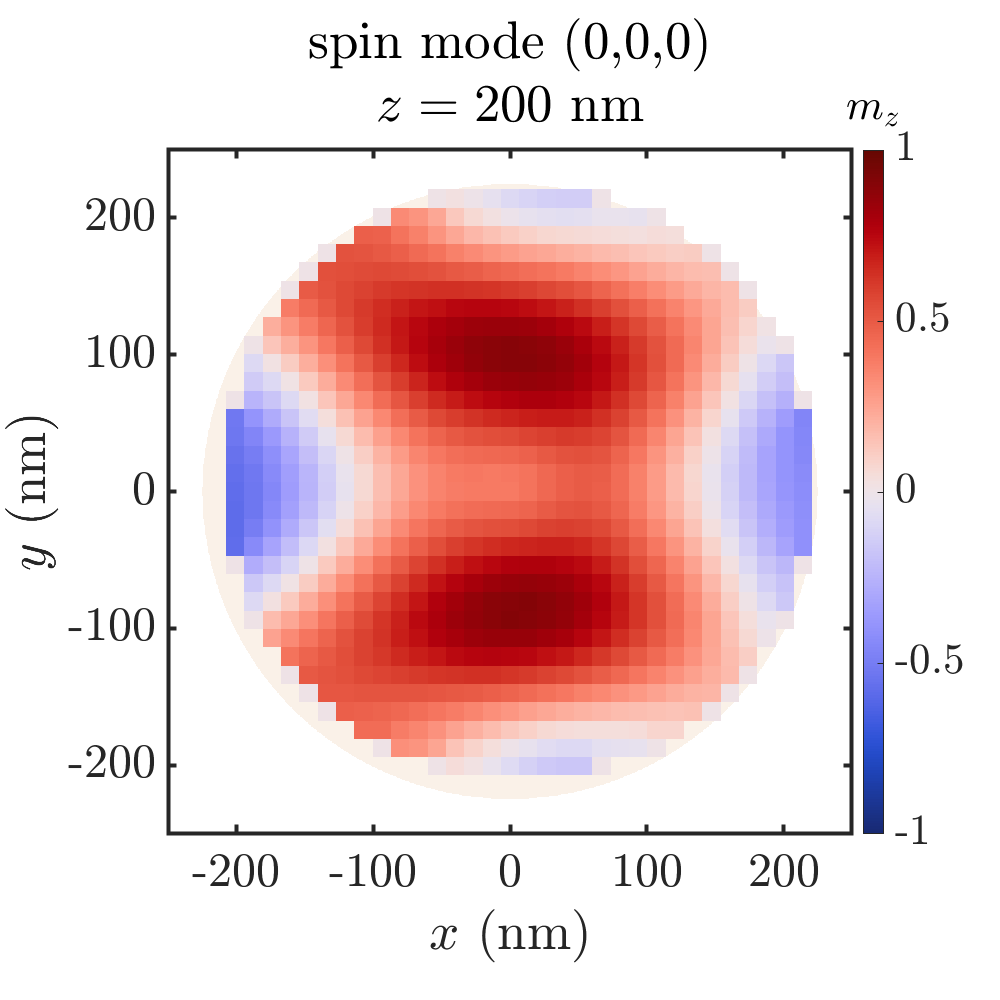}~~~
\includegraphics[width=0.32\linewidth]{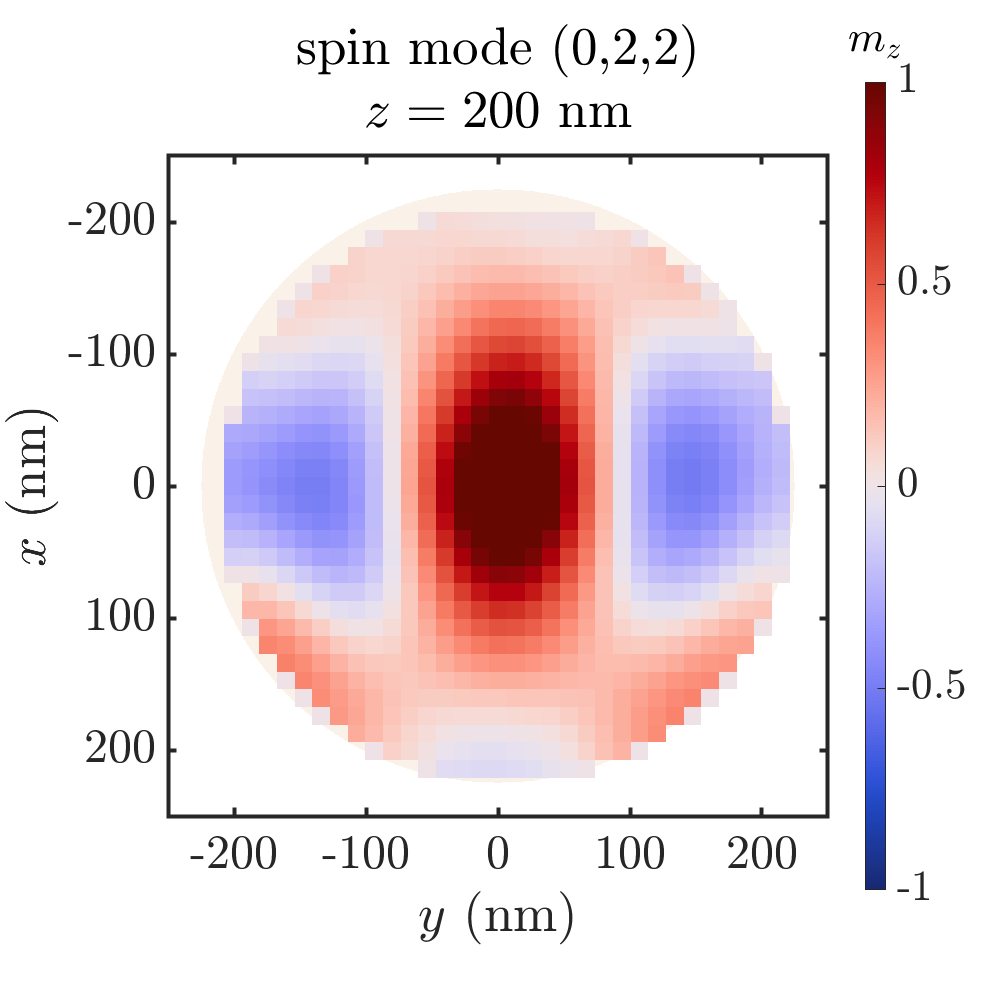}\\
\caption{Spin mode 3D profiles (top panel) and central cross-sections (bottom panel) of $m_z$ oscillating component of (a,d) edge-like standing spin mode $(2,0,0)$, (b,e) quasi-homogeneous mode $(0,0,0)$ and (c,f) standing spin mode $(0,2,2)$.}
\label{fig: Modes}
\end{figure*}

Figure~\ref{fig: Modes} shows the 3D profiles and central cross-sections of $m_z$ oscillating component of the standing spin modes launched in a nanocavity by a femtosecond laser pulse (see Fig.~\ref{fig: Precession}). One might see that all three standing spin modes excited in a nanocavity have quite complex profiles, and sign change is observed in all three spatial directions. This confirms that spin dynamics in the 3D optomagnonic nanocavity really have 3 degrees of freedom. Let us discuss the nature of the excited standing spin modes in more detail by analyzing the mode profiles and identifying the number of the nodes $n_x,~n_y,~n_z$ in the corresponding spatial directions.

The amplitude of the standing spin mode with the lowest frequency ($(2,0,0)$) is high near the nanocavity edges. Such a mode is similar to the edge spin mode of a disk~\cite{park2002spatially,barman2004imaging}; therefore, due to the demagnetization fields, it has a lower frequency than the other modes. At the same time, although the magnitude of $m_z$ decreases in the central part, it obviously has two nodes along the $x$ direction, while the phase in the $y$ and $z$ directions is quite homogeneous, so that it might be identified as $n_x=2, n_y=0, n_z=0$ edge-like standing spin mode.

The mode marked as $(0,0,0)$ is characterized by a quite homogeneous distribution of $m_z$ component inside a nanocavity. This makes it analogous to the fundamental mode of a nanocavity, except for some inhomogeneities at the nanocylinder edges. Therefore, this mode can be identified as the quasi-homogeneous mode of a nanocavity.

The mode with the highest frequency observed both in experiment and simulation is characterized by two nodes in  $y$ and $z$ spatial directions, so that it might be identified as $(0,2,2)$ standing spin mode. This is quite a remarkable result, since such a mode has a nearly zero average value of $\langle m_z \rangle=\frac{1}{V_\mathrm{cyl}}\int_{V_\mathrm{cyl}} m_z (\mathbf{r}) \mathrm{d}\mathbf{r}$. On the other hand, the spatial scale of mode profile inhomogeneity is about 150~nm. An ability to efficiently excite modes with such an amplitude distribution is due to a high localization of the optically generated effective magnetic field (Fig.~\ref{fig: Optical}b). Actually, the distance between the two neighbor nodes of this mode is close to the $H_\mathrm{IFE}$ spot size inside a cylinder, which makes it the highest-order mode excited in this configuration.
%\section*{Discussion}
To the best of our knowledge, we report the first experimental observation of optomagnetic excitation of standing spin modes quantized and inhomogeneous in all three spatial dimensions of a nanocavity. Due to the submicron size and consequent contributions of the exchange interaction and demagnetizing fields to spin dynamics, the frequencies of these modes differ significantly. Moreover, one may control the spin modes frequency splitting and magnitudes by the design of the nanocavity shape, for example, fabrication as nanocubes, nanoprisms, etc.
This opens wide opportunities for multichannel optomagnonic devices working simultaneously with several spin modes at different frequencies.

High confinement of the optically generated effective magnetic field makes it possible to excite standing spin modes with inhomogeneous and sign-changing profiles. Qualitatively, the amplitude $A$ of the certain spin mode with $m_z(\mathbf{r})$ distribution launched by the $H_\mathrm{exc}(\mathbf{r})$ magnetic field can be estimated as overlap integral $A \propto \int_V m_z(\mathbf{r}) H_\mathrm{exc}(\mathbf{r})$. Thus, if spin dynamics is launched by a quasi-homogeneous either optically-induced or microwave magnetic field, $H_\mathrm{exc}(\mathbf{r})\approx \mathrm{const}$, high-order modes with $n_j>0$ characterized by $\langle m_z \rangle \sim 0$ are excited with small amplitudes (see~\cite{chernov2020all}, for example).

In the studied nanocavity, the magnitudes of the excited standing spin modes are nearly the same (Fig.~\ref{fig: Precession}b), although two of them, $(2,0,0)$ and $(0,2,2)$ are characterized by very inhomogeneous profiles. One can numerically characterize the homogeneity of the spin mode profile with a value $\eta=\langle m_z \rangle/ \langle |m_z| \rangle$ with $\eta =1$ for the homogeneous modes and $\eta \rightarrow 0$ with the increase of the inhomogeneity. In our case, $\eta=0.9$ for quasi-homogeneous mode (0,0,0), and $\eta=0.5$ and $\eta=0.2$ for (0,2,2) and (2,0,0) modes, respectively. Excitation of such inhomogeneous standing spin modes is possible due to a high confinement of optically-generated inverse Faraday effect field.

In our case, optical localized resonance produces an effective magnetic field concentrated in the center of a nanocavity. At the same time, utilization of nanophotonic structures~\cite{ignatyeva2022all} with other types of optical modes and other optomagnetic effects~\cite{kalashnikova2015ultrafast}, such as inverse Coutton-Moutton effect, inverse transverse magneto-optical Kerr effect might open wide possibilities to perform launching of the other orders of spin modes via different shapes and phase profiles of the exciting magnetic field. Moreover, nanophotonics provides powerful means to easily tune the types of these excitations by changing the characteristics of the femtosecond pulses, such as their polarization and wavelength, which allow switching between different optical modes in one nanocavity.

%\section*{Conclusion}
We report the first experimental observation of the standing spin modes quantized and inhomogeneous in all three spatial dimensions. The recipe for their excitation involves femtosecond laser pulses illuminating an optomagnonic nanocavity at a wavelength of some optical resonance of the nanocavity. The optically induced field of the inverse Faraday effect is localized at the scale of a few hundred nanometers, thus providing a necessary point-like instant impact on spins and launching a set of different standing spin modes. Among them, the most interesting mode is the one with nodes in both lateral and longitudinal directions. Apart from that, its averaging along the whole nanocylinder is close to zero. It's excitation most vividly demonstrates the advantages of nanophotonics optomagnetism with respect to conventional means of spin wave excitation. The presented approach opens wide opportunities for multichannel optomagnonic devices tunably launching several spin modes at different frequencies.

\section*{acknowledgement}

This work was financially supported by the Russian Foundation for Basic Research, project No. 21-72-10020. The nanocylinders were fabricated at the Nano-fabrication facility, Nanocenter, University of Minnesota and the microfabrication facility (MFF) at Michigan Tech. University. ML and DK gratefully acknowledge support from the Michigan Tech Henes Center for Quantum Phenomena.


%\section*{Data availability}
%The data that support the plots within this paper and other findings of this study are available %from the corresponding authors Daria Ignatyeva (daria.ignatyeva@gmail.com)

\bibliography{bibliography}

\end{document}

\title{Supplemental Material for "Optical excitation of multiple standing spin modes in 3D optomagnonic nanocavities"}
\preprint{APS/123-QED}

\author{Daria O. Ignatyeva}
\email{daria.ignatyeva@gmail.com}
\affiliation{Russian Quantum Center, 121205 Moscow, Russia}
\affiliation{Faculty of Physics, M.V. Lomonosov Moscow State University, 119991 Moscow, Russia}

\author{Denis M. Krichevsky}
\affiliation{Russian Quantum Center, 121205 Moscow, Russia}
\affiliation{Moscow Institute of Physics and Technology, Moscow, Russia}

\author{Dolendra Karki}
\affiliation{Physics Department, Michigan Technological University, Houghton, Michigan, U.S.A.}

\author{Anton Kolosvetov}
\affiliation{Russian Quantum Center, 121205 Moscow, Russia}
\affiliation{Center for Photonics and 2D Materials, Moscow Institute of Physics and Technology (National Research University), Moscow, Russia}

\author{Polina E. Zimnyakova}
\affiliation{Russian Quantum Center, 121205 Moscow, Russia}
\affiliation{Moscow Institute of Physics and Technology, Moscow, Russia}

\author{Alexander N. Shaposhnikov}
\affiliation{Institute of Physics and Technology, V.I. Vernadsky Crimean Federal University, 295007 Simferopol, Russia}

\author{Vladimir N. Berzhansky}
\affiliation{Institute of Physics and Technology, V.I. Vernadsky Crimean Federal University, 295007 Simferopol, Russia}

\author{Miguel Levy}
\affiliation{Physics Department, Michigan Technological University, Houghton, Michigan, U.S.A.}

\author{Alexander I. Chernov}
\affiliation{Russian Quantum Center, 121205 Moscow, Russia}
\affiliation{Center for Photonics and 2D Materials, Moscow Institute of Physics and Technology (National Research University), Moscow, Russia}

\author{Vladimir I. Belotelov}
\affiliation{Faculty of Physics, M.V. Lomonosov Moscow State University, 119991 Moscow, Russia}
\affiliation{Russian Quantum Center, 121205 Moscow, Russia}
%\footnote{}

%\begin{tocentry}
%\includegraphics[width=3.3in]{TOC.png}
%\end{tocentry}

\date{\today}

\maketitle

\section{{Sample fabrication}}

First, bismuth-substituted iron garnet $\mathrm{Bi_{1.0}Lu_{0.5}Gd_{1.5}Fe_{4.2}Al_{0.8}O_{12}}$ (BIG) film of 515~nm thickness was deposited via magnetron sputtering on a $\mathrm{SiO_2}$ substrate. The  nanocylinders arrays were then fabricated via E-beam lithography (EBL) and Argon ions beam sputter etching. The nanocylinder patterns were written on 550 nm thick positive e-beam resist (ZEP-520A) by 100 KeV e-beams (VISTEC EBPG 5000+) exposed for a uniform dose of $\mathrm{140\mu C/cm^2}$ after proximity effect correction (PEC). Since ZEP is a non-conducting polymer resist, a 30 nm thick conducting gold layer coating was sputter deposited atop and wired to ground via conducting copper tape to avoid distortions of e-beam due to electrical charging during e-beam writing procedure. The exposed resist was developed in an amyl acetate solution after the removal of the gold layer in gold etchant solution first. The nanocylinder resist patterns were then printed on to BIG film by slow sputter etching at a rate of ~ 2.5 nm/minute in an Argon-ions-mill system (Intlvac Nanoquest, beam parameters: beam voltage- 200V, accelerating voltage – 24V, beam current -70 mA and plasma forward power-71 Watt). The sample stage was kept cooled at 6 deg. C throughout the etching duration to prevent hardening of the resist. Finally, the remaining thin layer of resist was removed by using resist remover N-methyl-2-pyrrolidine (NMP) solution heated at 80 deg. C. 

\section{{Time-resolved optomagnetic measurements}}

\begin{figure}[htbp]
\includegraphics[width=0.5\linewidth]{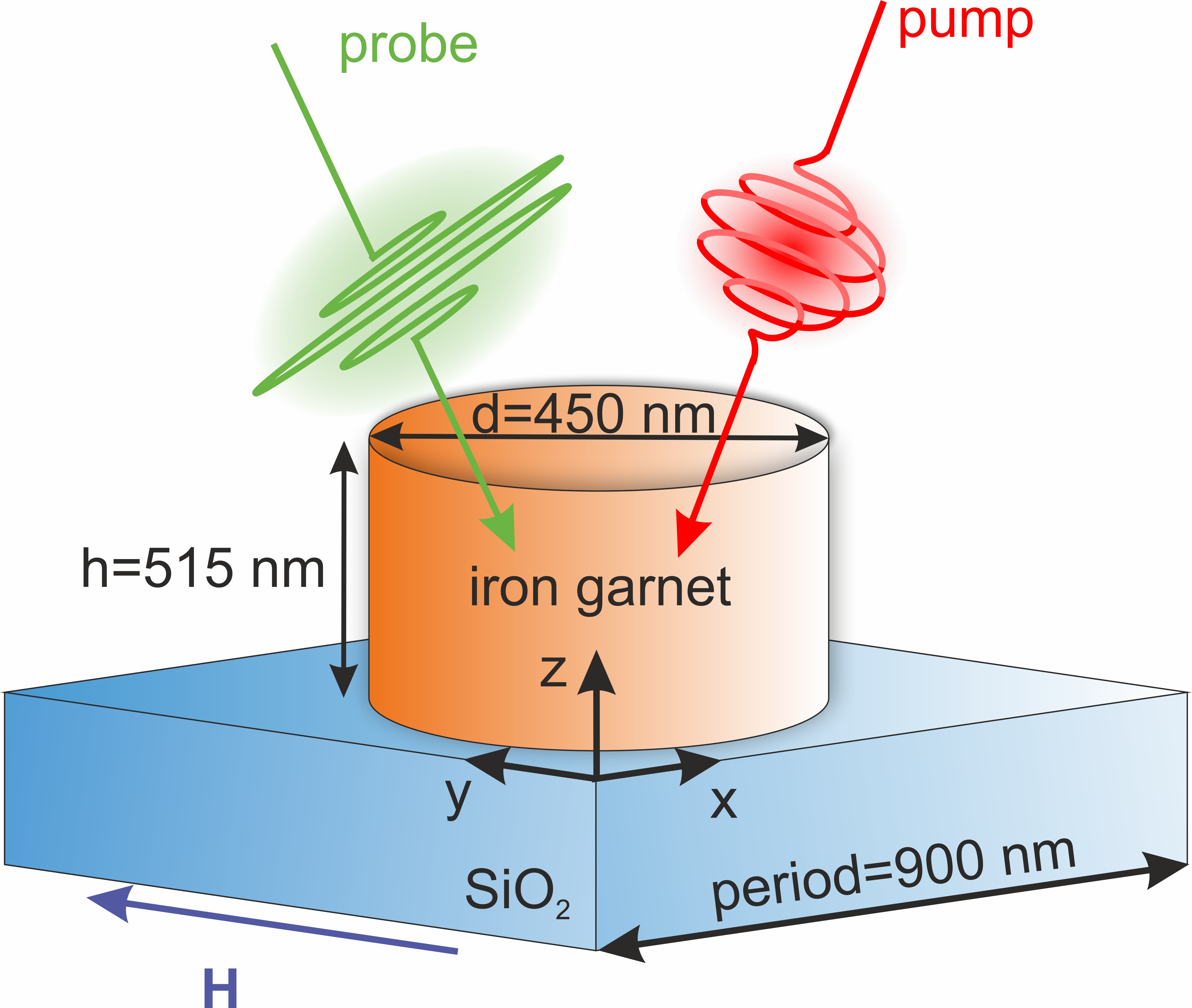}~~~
\caption{Configuration of magneto-optical pump-probe experiments: unit cell of the nanostructure under in-plane external magnetic field is excited by circularly polarized pump pulse and probed by linearly polarized light. }
\label{fig: config}
\end{figure}

Spin dynamics in the sample was experimentally measured using a time-resolved magnetooptical pump-probe scheme. Pump and probe pulses were generated by Avesta TOPOL parametric oscillator pumped by a Yb-doped Avesta TEMA laser ($80 \pm{5}$ MHz repetition rate, 180 fs pulse duration). Pump pulses of 730 nm wavelength were utilized to excite the system, and linearly polarized probe pulses of 525 nm were used to detect spin dynamics via the Faraday effect. Delayed pump pulses were modulated using a photoelastic modulator (Hinds Instruments PEM 100). The polarization changes of the probe pulse passed through the sample due to the Faraday effects were measured using an auto-balanced optical receiver (Nirvana 2007) in a lock-in detection scheme. The sample was placed in the in-plane field of an electromagnet. The scheme of the measurement configuration is presented in Figure~\ref{fig: config}.

\section{{Optical simulations}}

Electromagnetic simulations of optical modes inside the unit cell of the considered all-dielectric nanostructure were carried out by numerical
solution of Maxwell equations using the rigorous coupled-wave analysis (RCWA) approach~\cite{moharam1995formulation,li2003fourier}. Bloch boundary conditions were applied on the unit cell to simulate infinite array (along axis $x$ and $y$, see Figure~\ref{fig: config}). A plane wave excitation was utilized. Dispersion of BIG dielectric permittivity was taken into account~\cite{ignatyeva2020all}, so that $\varepsilon_{BIG}=7.2+i0.028$ for the wavelength of 730 nm. The refractive index of $\mathrm{SiO_2}$ was equal to 1.45.

\section{{Micromagnetic simulations}}

\begin{figure}[htbp]
\includegraphics[width=0.5\linewidth]{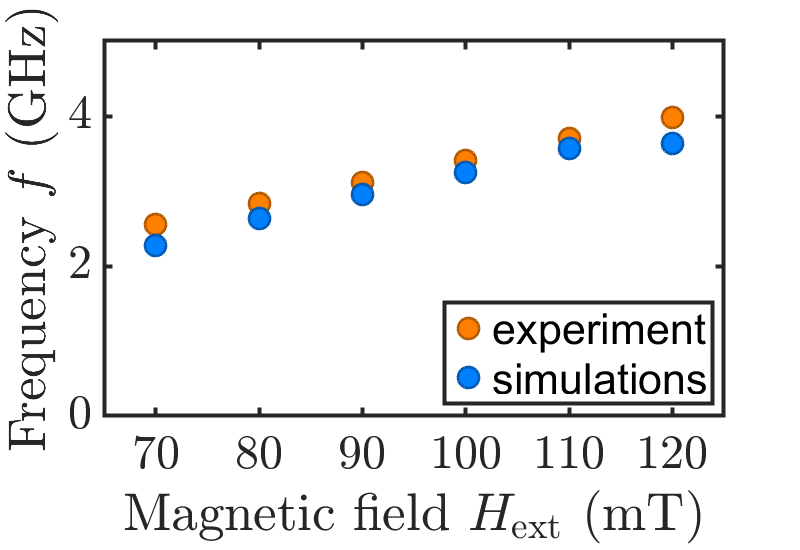}~~~
\caption{The resonant frequencies of magnetization precession obtained for the different values of the external field experimentally and numerically.}
\label{fig: f from H film}
\end{figure}

We built a numerical model of a bismuth-substituted iron-garnet nanocavity using the micromagnetic platform Mumax3 \cite{vansteenkiste2014design}. Micromagnetic modeling was first used to measure and verify the values of such material parameters as saturation magnetization \(M_{sat}\) and effective anisotropy \(K_u\). Spin dynamics was excited in a 550-nm thick film, subdivided into $128\times128\times1$~cells. The resonant frequencies of magnetization precession obtained for the different values of the external field show good agreement with the experimental results (Fig.~\ref{fig: f from H film}). The result of this estimation was \(M_s = 160\) kA/m, \(K_u = \) 2000 J/\(m^3\).

For spin-wave mode analysis, we created a micromagentic model of a single nanocylinder. The 900x900x620 nm simulation area was divided into 64x64x64 cells, with the dimensions of a single cell of 14x14x10 nm being smaller than the material exchange length of approximately 40 nm. We excited spin waves in a small $30 \times 30 \times 20$ nm area in the center of the cylinder by applying a time-varying magnetic field to the cells inside the area. 

We used a function $H_\mathrm{exc}\propto \sin(2 \pi f_0 t)/(2 \pi f_0 t)$ as a time-varying excitation magnetic field. The amplitude of the out-of-plane component of magnetization $m_z$ was measured with a 4~ps step in the central area of $\sim100$~nm radius inside a the nanocylinder. Precessional $m_z(x,y,z)$ part was obtained by subtracting the static magnetization $M_z(x,y,z)$ (see Fig.\ref{fig: Optical}). which mimicks the action of a probe femtosecond pulse. Such excitation with the cut-off frequency $f_0=10$~GHz allowed us to observe all of the modes obtained in experiment as the peaks of the Fourier transform of $m_z(t)$ temporal dependence at $f_j$ frequencies. (Fig.~\ref{fig: Precession}d). 

To study the profiles of the modes with these $f_j$ resonance frequencies, we used $H_\mathrm{exc}\propto \sin(2 \pi f_j t)$ time-varying excitation magnetic field applied to the central area of a nanocavity. The corresponding distributions of $m_z(x,y,z)$ were obtained after $N\sim15\gg1$ periods of the excitation magnetic field when the spin dynamics reached the stationary mode.

\bibliography{bibliography}